\documentclass[onecolumn,showpacs,preprintnumbers,amsmath,amssymb,aps,letterpaper]{revtex4}
\input{psfig.sty}

\begin{document}
\addtolength{\topmargin}{1.3cm}
\title{Observational H(z) Data as a complementarity to other cosmological probes}
\author{Hui Lin$^1$}
\author{Cheng Hao$^2$}
\author{Xiao Wang$^1$}
\author{Qiang Yuan$^3$}
\author{Ze-Long Yi$^4$}
\author{Tong-Jie Zhang$^{4,}$$^5$}
\email{tjzhang@bnu.edu.cn}
\author{Bao-Quan Wang$^6$}


\affiliation{$^1$Department of Statistics, School of Mathematical Sciences, Beijing Normal
University, Beijing, 100875, P.R.China}
\affiliation{$^2$Department of Applied Mathematics, School of Mathematical Sciences, 
Beijing Normal University, Beijing, 100875, P.R.China}
\affiliation{$^3$Key Laboratory of Particle Astrophysics, Institute
of High Energy Physics, Chinese Academy of Sciences, P.O.Box 918-3,
Beijing 100049, P.R.China}

\affiliation{$^4$Department
of Astronomy, Beijing Normal University, Beijing, 100875,
P.R.China}
\affiliation{$^5$Kavli Institute for theoretical Physics
China, Institute of Theoretical Physics, Chinese Academy of
Sciences(KITPC/ITP-CAS),P.O.Box 2735,Beijing
100875,P.R. China}
\affiliation{$^6$Department of Physics, Dezhou University, Dezhou, 253023, P. R. China}

\begin{abstract}
In this paper, we use a set of observational $H(z)$ data (OHD) to
constrain the $\Lambda$CDM cosmology. This data set can be derived
from the differential ages of the passively evolving galaxies.
Meanwhile, the $\mathcal {A}$-parameter, which describes the
Baryonic Acoustic Oscillation (BAO) peak, and the newly measured
value of the Cosmic Microwave Background (CMB) shift parameter
$\mathcal {R}$ are used to present combinational constraints on the
same cosmology. The combinational constraints favor an accelerating
flat universe
while the flat $\Lambda$CDM cosmology is also analyzed in the same
way. We obtain a result compatible with that by many other
independent cosmological observations. We find that the
observational $H(z)$ data set is a complementarity to
other cosmological probes.

\end{abstract}

\pacs{98.80.Cq; 98.80.Es; 04.50.+h; 95.36.+x} \maketitle

\section{Introduction}
Several observations, such as the type Ia Supernovae (SNe Ia)
\cite{Riess1998,Perlmutter1999}, Wilkinson Microwave Anisotropy
Probe (WMAP)
\cite{Spergel2003} and Sloan
Digital Sky Survey (SDSS)
\cite{Tegmark2004a,Tegmark2004b} support an accelerating expanding
universe. Many cosmological models such as the Quintessence
\cite{Alam2003}, the Braneworld cosmology\cite{DGP2000},
the Chaplygin Gas
\cite{Alcaniz2003} and the holographic dark energy models
\cite{Ke2005,Gong2005} are extensively
explored to explain the acceleration of the universe. One of the
most common candidates is composed of cold dark matter with the
equation of state $\omega=p/\rho=0$ and a cosmological constant
$\Lambda$ with $\omega=-1$. It is usually called the $\Lambda$CDM
cosmology
\cite{Carroll1992}, which is
characterized by
\begin{equation}
H^2(z)=H^2_0[\Omega_{\rm m}(1+z)^3+\Omega_{\Lambda}+\Omega_{\rm
k}(1+z)^2],\label{eq1}
\end{equation}
where $\Omega_{\rm m}$, $\Omega_{\Lambda}$ and $\Omega_{\rm k}$ are
proportion of the matter density, the cosmological constant and the
curvature term respectively, and $H_{\rm 0}=100h$ km s$^{-1}$
Mpc$^{-1}$ is the current value of the Hubble parameter. We can get
the relation $\Omega_{\rm k}=1-\Omega_{\rm m}-\Omega_{\Lambda}$ from
Eq.(\ref{eq1}) by setting $z=0$, and $\Omega_{\rm k}>0$,
$\Omega_{\rm k}=0$ and $\Omega_{\rm k}<0$ correspond to an open,
flat and closed universe respectively.

One important task is to constrain the cosmological parameters of
various cosmological models. Using the luminosity distance
measurements to a particular class of objects such as SNe Ia and
Gamma-Ray Bursts (GRBs) is the most frequently used method
\cite{Nesseris2004,Dai2004}. X-ray gas mass fraction of
galaxy clusters is also very popular
\cite{Zhu2004,Chang2006}. A different method based on the
lookback time measurements and the age of the universe has been used
to test the cosmological models
\cite{Capozziello2004,Dantas2006}. In addition, the size
of the Baryonic Acoustic Oscillation (BAO) peak detected in the
large-scale correlation function of luminous red galaxies from the
Sloan Digital Sky Survey (SDSS)
\cite{Eisenstein2005}, the Cosmic Microwave Background (CMB) data
obtained from the three-year WMAP estimate
\cite{Wang2006,Spergel2006} and the SNe Ia data
which are newly released as the first-year result of the planned
five-year Supernova Legacy Survey (SNLS) are widely used to
constrain cosmological models. Recently, one method based on the
observational $H(z)$ data (OHD for simplicity hereafter), which are
related to the differential ages of the oldest galaxies, has been
used to test cosmological models
\cite{Yi2007,Samushia2006,Wei2006,hy2007,Wu2007,Wei2007,Zhang2007.07,Danta2007,Zhang,Wei043009,Wu+,Wei+,Samushia+}.

In this paper, we first examine the non-flat $\Lambda$CDM cosmology
using the observational $H(z)$ data, which can be derived from the
derivative of redshift $z$ with respect to the cosmic time $t$
(i.e., ${\rm d} z/{\rm d}t$)
\cite{Jimenez2002}. Meanwhile, we do the combinational analyses
using observations of BAO and CMB.
We obtain a compatible result with many other independent
cosmological probes. We find that the observational $H(z)$ data are
effective for cosmological constraints.

We organize this paper as follows: In Sec.2, we briefly overview the
observational $H(z)$ data, the BAO data and the CMB data. In Sec.3,
we present the constraints on the $\Lambda$CDM model. In Sec.4, the
conclusions and more discussions are given.

\section{OHD, BAO and CMB as Combinational Tests}
\subsection{The Observational $H(z)$ Data}
The Hubble parameter $H(z)$ depends on the differential ages of the
universe in this form
\begin{equation}
H(z)=-\frac{1}{1+z}\frac{{\rm d}z}{{\rm d}t},\label{eq2}
\end{equation}
which provides a direct measurement for $H(z)$ through a
determination of ${\rm d}z/{\rm d}t$. In the work of Jimenez et al.
\cite{Jimenez2003}, they demonstrated the feasibility of the
method by applying it to a $z \sim 0$ sample. In particular, SDSS
was used to determine $H_0$ and it was showed that its value was in
good agreement with other independent methods. With the availability
of new galaxy surveys, it becomes possible to determine $H(z)$ at
$z>0$ . By using the differential ages of passively evolving
galaxies determined from the Gemini Deep Deep Survey (GDDS)
\cite{Abraham2004} and archival data
\cite{Treu2001,Treu2002,Nolan2003a,Nolan2003b}, Simon et al. (2005)
\cite{Simon2005} derived a set of observational $H(z)$ data.

The data set was derived from the absolute ages of 32 passively
evolving galaxies with high-quality spectroscopy. Synthetic stellar
population models were used to constrain the ages of the oldest
stars in the galaxy (after marginalising over the metallicity and
star formation history), similar to the work of Jimenez et al.
\cite{Jimenez2003}. In order to estimate the differential
ages, these galaxies were further divided into three subsamples. The
first subsample was composed of 10 field early-type galaxies, after
discarding galaxies for which the spectral fit indicated an extended
star formation. The ages of this sample were derived using the SPEED
models
\cite{Jimenez2004}. The second
subsample was composed of 20 old passive galaxies from GDDS. They
calculated the absolute ages using the SPEED models too, obtaining
harmonious results with the GDDS collaboration which estimated the
ages using different models.
The third subsample consisted of two red radio galaxies 53W091 and
53W069
\cite{Dunlop1996,Spinrad1997,Nolan2003b}. After grouping together
all the galaxies within $\Delta z=0.03$ and excluding unperfect
ones, they computed differential ages only for those bins within
$0.1\leq \Delta z\leq 0.15$.
The interval $\Delta z=0.03$ is set small in order to avoid
incorporating galaxies that have already evolved in age, but large
enough for the sparse sample to have more than one galaxy in most of
the bins.
The lower limit is imposed so that the age evolution between the two
bins is larger than the error in the age determination. This
provides a robust determination of $dz/dt$. The differential ages
are less sensitive to systematics errors than absolute ages
\cite{Jimenez2004}. Then a set of differential ages
$dz/dt$, equivalently $H(z)$, was obtained.
As $z$ has a relatively wide range, $0.1<z<1.8$, these data are
expected to provide a more full-scale description of the dynamical
evolution of our universe. But what a pity, the data amount is not
sufficient enough and the corresponding errors are quite large
\cite{Samushia2006,Wei2006}.


These observational $H(z)$ data have been used to constrain the dark
energy potential and its redshift dependence by Simon et al.
\cite{Simon2005}. Using this data set, one can constrain various
cosmological models too. Yi \& Zhang
\cite{Yi2007} first used
them to analyze the holography-inspired dark energy models in which
the parameter $c$ plays a key role. The cases with $c=0.6, 1.0, 1.4$
and setting $c$ free are discussed in detail. The results are
consistent with some other independent cosmological tests. Samushia
\& Ratra
\cite{Samushia2006} used the data set to constrain
the parameters of $\Lambda$CDM, XCDM and $\phi$CDM models. Wei \&
Zhang
\cite{Wei2006} compared a series of other cosmological
models with interaction between dark matter and dark energy. And
they find that the best models have an oscillating feature for both
the Hubble parameter and the equation of state.





\subsection{The BAO Data}
The acoustic peaks in the CMB anisotropy power spectrum has been
found efficient to constrain cosmological parameters
\cite{Spergel2003}. As the acoustic oscillations in the
relativistic plasma of the early universe will also be imprinted on
to the late-time power spectrum of the non-relativistic matter
\cite{Eisenstein1998}, the acoustic
signatures in the large-scale clustering of galaxies yield
additional tests for cosmology.

Using a large spectroscopic sample of 46748 luminous red galaxies
covering 3816 square degrees out to $z=0.47$ from the SDSS,
Eisenstein et al.
\cite{Eisenstein2005} successfully found
the peaks, described by the model-independent $\mathcal
{A}$-parameter which is independent on $H_0$,
\begin{equation}
\mathcal {A}=\frac{\sqrt{\Omega_{\rm
m}}}{z_1}[\frac{z_1}{E(z_1)}\frac{1}{|\Omega_{\rm k}|}{\rm
sinn}^2(\sqrt{|\Omega_{\rm k}|}\mathcal {F}(z_1))
]^{1/3},\label{eq3}
\end{equation}
where $E(z)=H(z)/H_0$, $z_1=0.35$ is the redshift at which the
acoustic scale has been measured, the function ${\rm sinn(x)}$ is
defined as
\begin{equation}
{\rm sinn(x)}\equiv\left\{
\begin{array}{lll}
{\rm sinh(x)} & {\rm if}\ \Omega_{\rm k}>0;\\
{\rm x} & {\rm if}\ \Omega_{\rm k}=0;\\
{\rm sin(x)} & {\rm if}\ \Omega_{\rm k}<0,
\end{array}\right.\label{eq4}
\end{equation}
and the function $\mathcal {F}(z)$ is defined as
\begin{equation}
\mathcal {F}(z)\equiv\int_0^z\frac{dz}{E(z)}.\label{eq5}
\end{equation}
Eisenstein et al.
\cite{Eisenstein2005} suggested the
measured value of the $\mathcal {A}$-parameter as $\mathcal
{A}=0.469\pm0.017$.

\subsection{The CMB Data}
The shift parameter $\mathcal {R}$ is perhaps the most
model-independent parameter which can be derived from CMB data and
it does not depend on $H_0$. It is defined as
\cite{Bond1997,Odman2003}
\begin{equation}
\mathcal {R}=\frac{\sqrt{\Omega_{\rm m}}}{\sqrt{|\Omega_{\rm
k}|}}{\rm sinn}[\sqrt{|\Omega_{\rm k}|}\mathcal {F}(z_{\rm
r})],\label{eq6}
\end{equation}
where $z_{\rm r}=1089$ is the redshift of recombination. From the
three-year result of WMAP
\cite{Spergel2006}, Wang \& Mukherjee
\cite{Wang2006} estimated the CMB shift parameter $\mathcal {R}$ and
showed that its measured value is mostly independent on assumptions
about dark energy. The observational result is suggested as
$\mathcal {R}=1.70\pm0.03$
\cite{Wang2006}.

BAO and CMB have been widely used to do combinational constraints on
the cosmological parameters. In the work of Guo et al.
\cite{Guo2006}, BAO was used to constrain the
Dvali-Gabadadze-Porrati (DGP) braneworld cosmology and a closed
universe is strongly favored. Wu \& Yu
\cite{Wu2006} combined
BAO with CMB to determine parameters of a dark energy model with
$\omega=\omega_0/[1+b{\rm ln}(1+z)]^2$. They suggested that a
varying dark energy model and a crossing with $\omega=-1$ are
favored and the current value of $\omega$ is very likely less than
-1. Pires et al.
\cite{Pires2006} combined BAO and the
lookback time data to make a joint statistic analysis for the DGP
braneworld cosmology and they suggested a closed universe. Wang \&
Mukherjee
\cite{Wang2006} used the $\mathcal {R}$ parameter,
combing several other cosmological probes including BAO, to derive
model-independent constraints on the dark energy density and the
Hubble parameter.

\section{Constraints on the $\Lambda$CDM Cosmology}

\begin{figure*}
\centerline{\psfig{figure=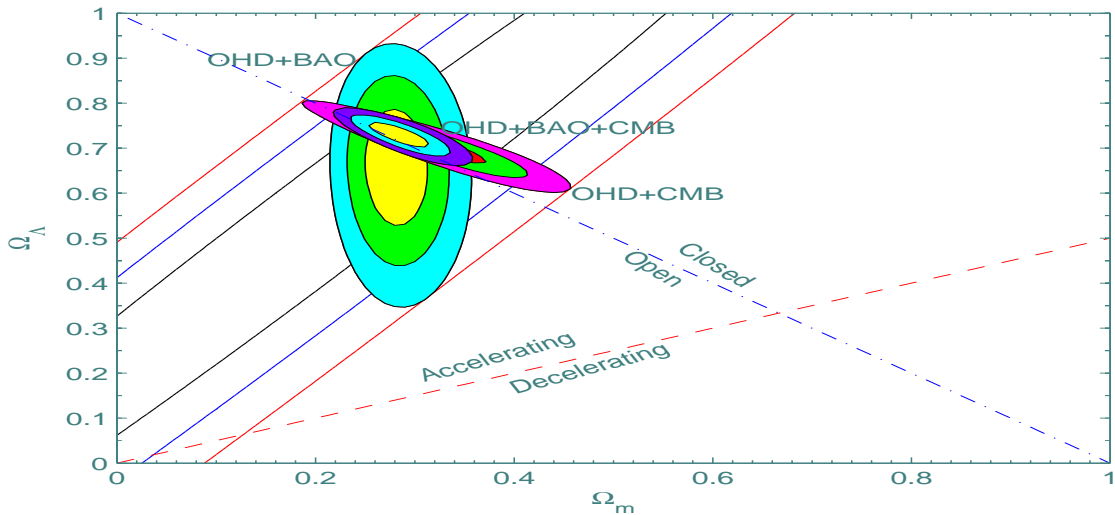,width=6.7in,height=2.9in,angle=0}}{\hskip
0.1in} \caption {Confidence regions in the $\Omega_{\rm
m}-\Omega_{\Lambda}$ plane for a non-flat $\Lambda$CDM universe, for
different observational data sets as labeled in the figure. The
shaded regions from inner to outer stand for confidence levels at
68.3\%, 95.4\% and 99.7\% respectively. For a comparison, confidence
regions for considering only the observational $H(z)$ data are
plotted with solid lines. And the solid lines from inner to outer
correspond to confidence levels at 68.3\%, 95.4\% and 99.7\%
respectively. The dash-dot straight line represents a flat universe
which satisfies $\Omega_{\rm k}=0$, i.e., $\Omega_{\rm
m}+\Omega_{\Lambda}=1$. The dashed straight line is the critical
line for an accelerating and a decelerating universe. }\label{fig1}
\end{figure*}

In this paper, we combine the observational $H(z)$ data with BAO and
CMB to make a constraint on the $\Lambda$CDM cosmology. The best-fit
parameters can be determined through the $\chi^2$ minimization
method.
For the non-flat $\Lambda$CDM cosmology, we assume a prior of
$h=0.72\pm0.08$ suggested by the Hubble Space Telescope (HST) Key
Project
\cite{Freedmann2001}. Thus we have only two free parameters, i.e.,
$\Omega_{\rm m}$ and $\Omega_{\Lambda}$. We get the fitting results
$\Omega_{\rm m}=0.19\pm0.34$ and $\Omega_{\Lambda}=0.52\pm0.57$,
with the $\chi^2$-value per degree of freedom $\chi^2_{\rm
min}/d.o.f=8.94/7$. The best-fit results suggest $\Omega_{\rm
k}=0.29$. We plot the confidence regions in the $\Omega_{\rm
m}-\Omega_{\Lambda}$ plane in Fig.\ref{fig1}. It seems that this
constraint is quite weak and requires deeper discussions such as
combinational analysis with other cosmological probes.

In order to further explore the role of the observational $H(z)$
data, we study the combinational constraints from the other
cosmological probes. If we combine OHD and BAO, we get $\Omega_{\rm
m}=0.28\pm0.02$ and $\Omega_{\Lambda}=0.66\pm0.09$, with
$\chi^2_{\rm min}/d.o.f=9.01/8$. The best-fit results correspond to
a universe with $\Omega_{\rm k}=-0.04$.
We also plot the confidence regions in the $\Omega_{\rm
m}-\Omega_{\Lambda}$ plane in Fig.\ref{fig1}. It is clear that the
constraint from OHD+BAO is much more restrict than using only OHD.
We also find that $\Omega_{\rm m}$ is constrained more effectively
than $\Omega_{\Lambda}$.

By combing OHD and CMB, we get $\Omega_{\rm m}=0.31\pm0.04$ and
$\Omega_{\Lambda}=0.71\pm0.03$, with $\chi^2_{\rm
min}/d.o.f=9.06/8$. The best-fit values correspond to a universe
with $\Omega_{\rm k}=-0.02$. The fitting results are not too far
from the result for combining OHD+BAO, but they suggest a universe
closer to being flat. We plot the confidence regions in the
$\Omega_{\rm m}-\Omega_{\Lambda}$ plane in Fig.\ref{fig1} too, and
the parameters are constrained more strictly than that using OHD and
OHD+BAO, especially $\Omega_{\Lambda}$. And the tendency of the
confidence regions keeps along with the dash-dot straight line which
corresponds to a flat universe. Note Fig.4 in the work of Su et al.
\cite{Su2006} for more comparisons with results from GRBs.


If we combine OHD, BAO and CMB, we get $\Omega_{\rm m}=0.28\pm0.02$
and $\Omega_{\Lambda}=0.73\pm0.02$, with $\chi^2_{\rm
min}/d.o.f=9.74/9$. This case corresponds to a universe with
$\Omega_{\rm k}\simeq-0.01$, which is a flat universe at a
confidence level of 68.3\%. Confidence regions in the $\Omega_{\rm
m}-\Omega_{\Lambda}$ plane are plotted in Fig.\ref{fig1}. Both
$\Omega_{\rm m}$ and $\Omega_{\Lambda}$ are constrained more
strictly. Clearly, all points within the confidence region at 99.7\%
confidence level are near the dash-dot straight line which stands
for a flat universe. And our confidence regions are very similar to
those from BAO+CMB+GRBs
\cite{Su2006}. All the fitting results discussed above are listed in
Table \ref{table1}.

\begin{table}
\caption{Fitting results for a non-flat universe }
\begin{ruledtabular}
\begin{tabular}{lcrlccc}
Test     & $\Omega_{\rm m}$  & $\Omega_{\rm \Lambda}$  & $\Omega_{\rm k}$ & $\chi^2_{\rm min}/d.o.f$  & \\
\hline
OHD         & $0.19\pm0.34$ & $0.52\pm0.57$  &0.29&8.94/7\\
  OHD+BAO   &   $0.28\pm0.02$ & $0.66\pm0.09$   &-0.04&9.01/8\\
  OHD+CMB         &   $0.31\pm0.04$ & $0.71\pm0.03$ &-0.02&9.06/8\\
  OHD+BAO+CMB   & $0.28\pm0.02$ & $0.73\pm0.02$  &-0.01& 9.74/9\\
\end{tabular}
\end{ruledtabular}\label{table1}
\end{table}

\begin{figure*}
\centerline{\psfig{figure=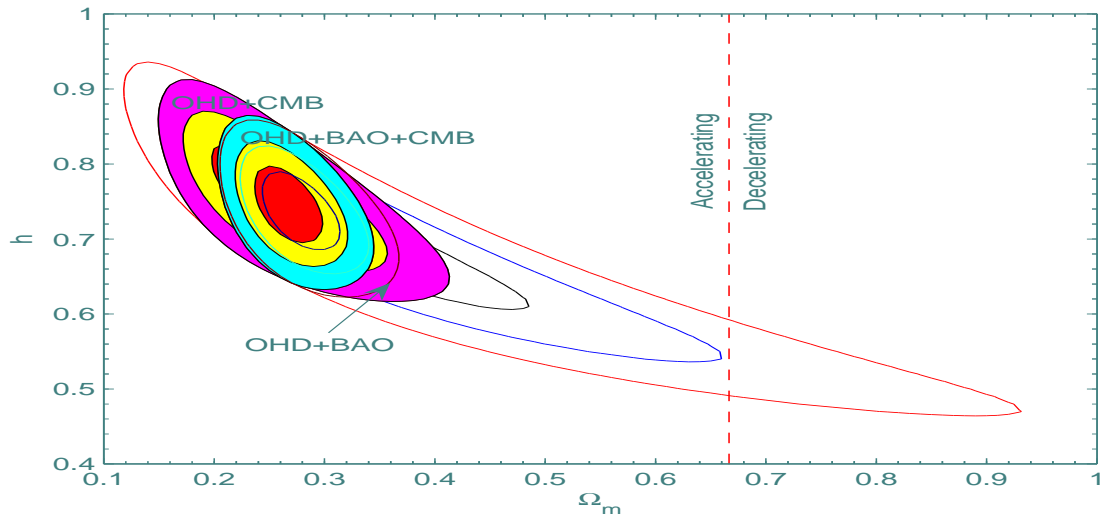,width=6.7in,height=2.9in,angle=0}}{\hskip
0.1in} \caption {Confidence regions in the $\Omega_{\rm m}-h$ plane
for a flat $\Lambda$CDM universe. The most outer regions correspond
to OHD, from inner to outer for confidence levels at 68.3\%, 95.4\%
and 99.7\% respectively. Regions for different observational data
sets are labeled in the figure, from inner to outer for confidence
levels at 68.3\%, 95.4\% and 99.7\% respectively. The dashed
straight line is the critical line for an accelerating and a
decelerating universe.
}\label{fig2}
\end{figure*}

For a comparison, we present a constraint on the flat cosmology.
Thus only one free parameter needs to be fitted if the prior of $h$
is taken too. All the fitting results are listed in Table
\ref{table2}. All the tests suggest $\Omega_{\rm m}\simeq0.28$ which
is consistent with WMAP which suggested $\Omega_{\rm m}=0.27\pm0.04$
\cite{Spergel2003}. In addition, we discuss
the flat $\Lambda$CDM cosmology if $h$ is set free. All the fitting
results are listed in Table \ref{table3}. The information on
$\Omega_{\rm m}$ is consistent with that of WMAP
\cite{Spergel2003} and the value of $h$ is roughly concordant
with the prior we have taken in the above discussions. We plot the
confidence regions in the $\Omega_{\rm m}-h$ plane in
Fig.\ref{fig2}.

\begin{table}
\caption{Fitting results for a flat universe with a prior of $h$ }
\begin{ruledtabular}
\begin{tabular}{lcrlccc}
Test     & $\Omega_{\rm m}$  & $\Omega_{\rm \Lambda}$  & $\chi^2_{\rm min}/d.o.f$  & \\
\hline
OHD         &  $0.30\pm0.04$ & $0.70$ &9.04/8\\
  OHD+BAO   &  $0.28\pm0.02$ & $0.72$& 9.47/9\\
  OHD+CMB         &  $0.27\pm0.03$ & $0.73$ &10.97/9\\
  OHD+BAO+CMB   &  $0.27\pm0.02$ & $0.73$ &10.97/10\\
\end{tabular}
\end{ruledtabular}\label{table2}
\end{table}

\begin{table}
\caption{Fitting results for a flat universe, setting $h$ free }
\begin{ruledtabular}
\begin{tabular}{lcrlccc}
Test    & $h$  & $\Omega_m$  & $\Omega_{\Lambda}$  & $\chi^2_{\rm min}/d.o.f$  & \\
\hline
 OHD         &  $0.71\pm0.07$ &$0.31\pm0.09$ & $0.69$ & 9.02/7 \\
  OHD+BAO     &  $0.74\pm0.03$ &$0.28\pm0.02$ & $0.72$ & 9.23/8  \\
  OHD+CMB     &  $0.76\pm0.04$ &$0.25\pm0.04$ & $0.75$ & 10.03/8 \\
  OHD+BAO+CMB &  $0.75\pm0.03$ &$0.27\pm0.02$ & $0.73$ & 10.39/9 \\
\end{tabular}
\end{ruledtabular}\label{table3}
\end{table}

\section{Ihe Roles of OHD and SNe Ia data in Cosmological
Constraints}
The purpose of this section is to examine the role of OHD and SNe Ia
data in the constraints on cosmological parameters. Thus we just
consider the $\Lambda$CDM cosmological model with a curvature term.
The likelihood for the cosmological parameters can be determined
from a $\chi^2(h,\Omega_{\rm M},\Omega_\Lambda)$ statistic. We marginalize
the likelihood functions over $h$ by integrating the probability
density $P\propto e^{-\chi^2/2}$ to obtain the fitting results and
the confidence regions in the $\Omega_{\rm M}$-$\Omega_{\Lambda}$
plane.

\begin{figure}
\centerline{\psfig{figure=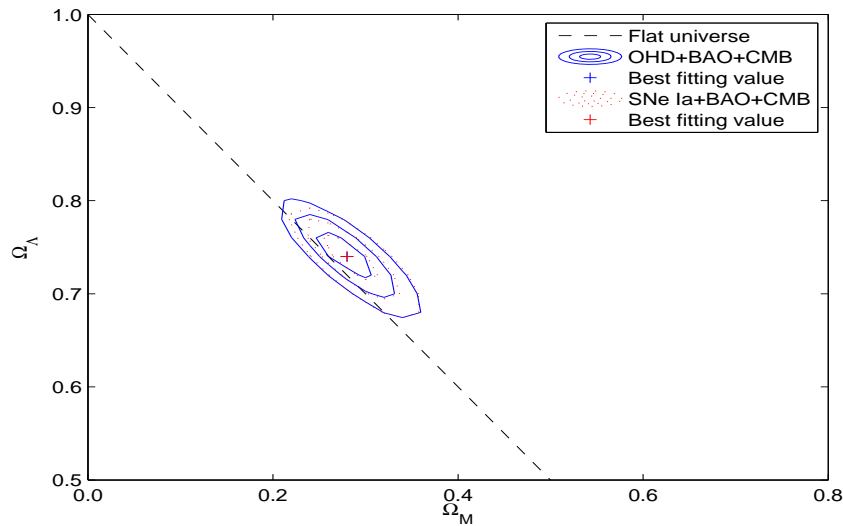,width=5.truein,height=3.truein,angle=0}}
\caption {Confidence regions at 68.3\%, 95.4\% and 99.7\%
levels from inner to outer respectively in the $\Omega_{\rm
M}-\Omega_{\Lambda}$ plane for a non-flat $\Lambda$CDM universe.
The dotted lines correspond to the results using SNe Ia+BAO+CMB,
while solid lines for OHD+BAO+CMB. The crosses in the
center of confidence regions indicate the best-fit values respectively.
The dash straight line represents a flat universe with $\Omega_{\rm k}=0$.
}
\label{fig3}
\end{figure}

\begin{figure}
\centerline{\psfig{figure=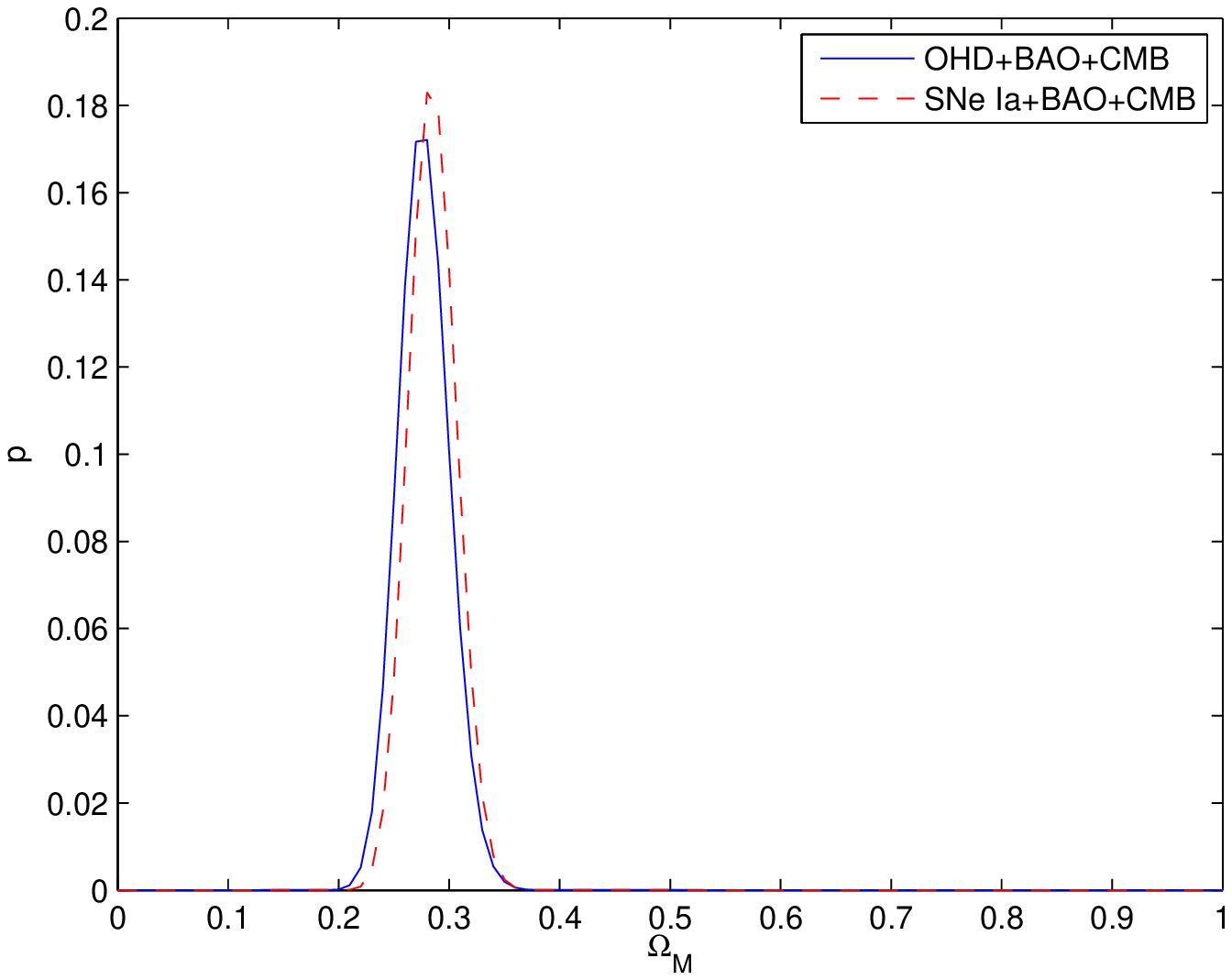,width=5.truein,height=3.truein,angle=0}}
\centerline{\psfig{figure=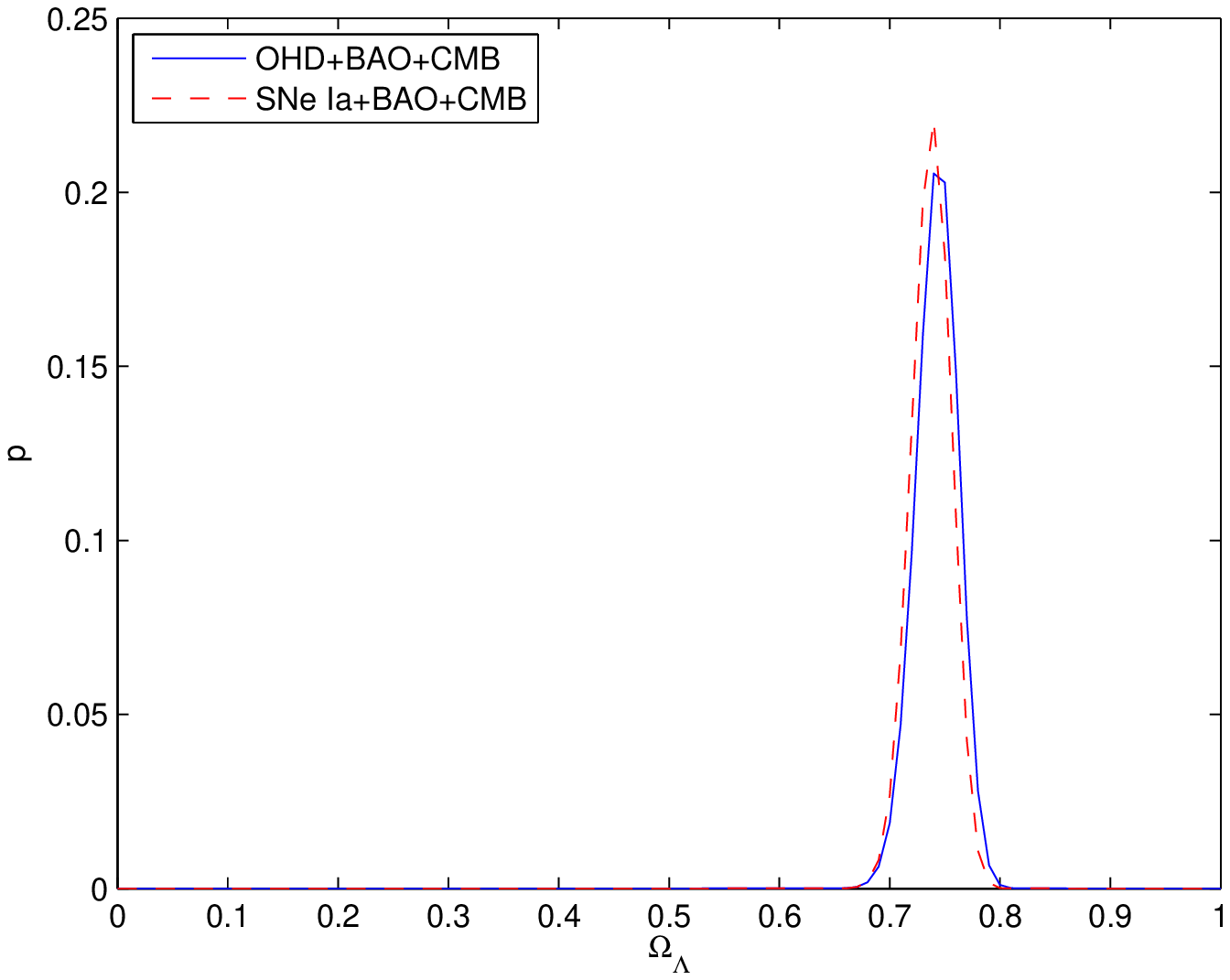,width=5.truein,height=3.truein,angle=0}}
\caption{The one-dimensional probability distribution function (PDF)
$p$ for selections of parameters $\Omega_{\rm M}$ (\emph{Top panel})
and $\Omega_{\Lambda}$(\emph{Bottom panel}) for the combination analysis
OHD+BAO+CMB (solid) and SNe Ia+BAO+CMB (dotted) respectively.}
\label{fig4}
\end{figure}

We further do the joint constraint using several data sets including
BAO and CMB. In order to avoid double-counting the constraints
from CMB observation, we do not employ the prior of $h$.
In Fig.\ref{fig3}, we show the combination results of OHD+BAO+CMB
(solid lines) and SNe Ia+BAO+CMB (dotted lines) respectively. 
For the combined constraint, we get $\Omega_{\rm M}=0.275$,
$\Omega_{\Lambda}=0.74$ for
OHD+BAO+CMB and $\Omega_{\rm M}=0.28$, $\Omega_{\Lambda}=0.74$ for SNe Ia+BAO+CMB respectively. 
The best-fit
results for OHD+BAO+CMB are almost identical to those for SNe
Ia+BAO+CMB combination, and
are consistent with that from WMAP five year's results \cite{wmap5}.
From Fig.\ref{fig3},
it is shown that there are slight differences between the confidence
regions at 68.3\%, 95.4\% and 99.7\% levels for OHD+BAO+CMB and
SNe Ia+BAO+CMB. The one-dimensional probability distribution
functions (PDF) $p$ for selections of parameters $\Omega_{\rm M}$
and $\Omega_{\Lambda}$ for the combination data analysis are shown
in Fig.\ref{fig4}. It is also shown that there are only slight
discrepancies between OHD+BAO+CMB and SNe Ia+BAO+CMB combination
data on both PDF of $\Omega_{\rm M}$ and $\Omega_{\Lambda}$.
Therefore, we find that the OHD play an almost same role as SNe Ia
for the joint constraints. In the recent work by Carvalho et
al.\cite{Carvalho2008}, the very similar conclusions are also drawn
in the test of $f(R)$ cosmology with the same data as our work.

\section{Conclusions and Discussions}
Recent observations have provided many robust tools to analyze the
dynamical behavior of the universe. Most of them are based on
distance measurements, such as SNe Ia. It is also important to use
other different probes to set bounds on the cosmological parameters.
In this paper, we have followed this direction and used the
observational $H(z)$ data from the differential ages of the
passively evolving galaxies to constrain the $\Lambda$CDM cosmology,
combining BAO and CMB.
For the non-flat case, the value of $\chi^2_{\rm min}/d.o.f$ from
OHD+BAO+CMB is the smallest while that from OHD is the largest. In
other words, OHD fails to provide a more restrict constraint on this
model. It is mainly due to its lackness in quality and big
observational errors which can be seen in Fig.\ref{fig1} of Yi and
Zhang
\cite{Yi2007}. However, most of the combinational results
suggest a universe with small absolute values of $\Omega_{\rm k}$,
i.e., close to being flat, which can be easily found in
Fig.\ref{fig1}. This is consistent with most observations that
support a flat universe
\cite{Spergel2003}.

For a flat universe with a prior, the value of $\chi^2_{\rm
min}/d.o.f$ from OHD+BAO is the smallest while that from OHD+CMB is
the largest. If we set $h$ free, the value of $\chi^2_{\rm
min}/d.o.f$ from OHD+CMB is the smallest while that from OHD+BAO is
the largest. From Fig.\ref{fig2}, we can see clearly that most of
the confidence regions are overlapped, in agreement with each other
very well. Meanwhile, most of the fitting results suggest an
accelerating expanding universe at 99.7\% confidence level.

From the above comparison and previous works \cite{Yi2007,Samushia2006}, 
we find that our results from the
observational $H(z)$ data are believable and the observational
$H(z)$ data can be seen as a complementarity to other
cosmological probes.


\section{Acknowledgments}
We are very grateful to the anonymous referee for his valuable
comments that greatly improve this paper. 
Ze-Long Yi would like to thank Yuan Qiang and Jie Zhou for their
valuable suggestions. This work was supported by the National
Science Foundation of China (Grants No.10473002, 10533010), 2009CB24901 
and the Scientific Research Foundation for the Returned 
Overseas Chinese Scholars, State Education Ministry.


\end{document}